# Observation of room-temperature intrinsic antiferroelectricity in van der Waals $NbOCl_2$

JinKai Huo,[1] Xingyu Chen,[1] Gaolei Zhao,[1] Xuan Chen,[1] Yuchen Meng,[1] Xu Han,[1] Juhe Liu,[1] Egor Kolesov,[2] Konstantin Kozadaev,[3] Alexei Tolstik,[4] Andrei Zheltkovich,[5] Andrey Novitsky,[6] Shenghuang Lin,[7] Changsen Sun,[1] and Dawei Li[1,*]

[1] School of Optoelectronic Engineering and Instrumentation Science, Dalian University of Technology, Dalian 116024, China

[2] Solid State Physics and Nanotechnology Department, Belarusian State University, Minsk, 220030, Belarus

[3] National Academy of Sciences of Belarus, Minsk, 220072, Belarus

[4] Laser Physics and Spectroscopy Department, Belarusian State University, Minsk, 220030, Belarus

[5] Theoretical and Applied Mechanics Department, Brest State Technical University, Brest, 224017, Belarus

[6] Physical Optics and Applied Informatics Department, Belarusian State University, Minsk, 220030, Belarus

[7] Dongguan Institute of Materials Science and Technology, Chinese Academy of Sciences, Dongguan 523808, China

* Correspondence to: dwli@dlut.edu.cn

**Abstract**

In recent years, two-dimensional (2D) ferroelectricity has been extensively investigated across a wide range of 2D materials, while intrinsic 2D antiferroelectricity remains exceptionally rare, limiting its implementation in next-generation photonic and electronic devices. Here, we report the observation of room-temperature-stable antiferroelectricity in the layered transition metal oxyhalide $NbOCl_2$, and demonstrate a reversible, stimuli-driven antiferroelectric-to-ferroelectric (AFE-to-FE) phase transition in this material. Second-harmonic generation (SHG) mapping reveals a well-defined stripe-like spatial distribution of SHG intensity across as-grown $NbOCl_2$ flakes spanning thickness from few-layer to bulk-like configurations. This distinctive SHG signature originates from the existence of antiferroelectric phase and the coexistence of antiferroelectric and ferroelectric phases within individual crystals, as evidenced by a combined incident-angle-dependent SHG measurement and Raman spectroscopy. By integrating $NbOCl_2$ with one-dimensional silver nanowires, we achieve spatially selective, strain-mediated AFE-to-FE phase transition. Moreover, an ultralow electric field (~20 kV/cm) induces a reversible SHG response, consistent with switching from the centrosymmetric antiferroelectric state (SHG-silent) to the non-centrosymmetric ferroelectric state (SHG-active). This study not only expands the family of room-temperature 2D antiferroelectrics but also establishes a versatile platform for dynamic manipulation of nonlinear light, paving the way toward electric-field-programmable nonlinear photonics and optical information processing.

## 1 Introduction

Conventional ferroelectrics, such as $PbTiO_3$ and $BiFeO_3$, possess spontaneous polarization that can be reversibly switched by an external electric field, enabling a broad range of functionalities of nonvolatile memories,[1-3] logic circuits,[4, 5] sensors,[5, 6] and optoelectronic devices.[7] However, most of these ferroelectric materials cannot maintain stable ferroelectricity in ultrathin films, limiting their implementation in ultimately scaled devices.[8, 9] In recent years, two-dimensional (2D) van der Waals (vdW) ferroelectrics have attracted much attention due to their intrinsic scaling advantages.[10, 11] So far, numerous 2D ferroelectrics have been discovered, such as $CuInP_2S_6$ with out-of-plane polarization,[12, 13] α-$In_2Se_3$ with intercorrelated out-of-plane and in-plane polarization,[14, 15] and $NbOX_2$ (X = Cl, I, Br) with in-plane polarization.[16-18] As a new family of 2D ferroelectrics, $NbOX_2$ exhibits room-temperature ferroelectricity, high spontaneous polarization,[16, 17] weak interlayer coupling,[19, 20] pronounced nonlinear optical responses,[19, 21, 22] and in-plane anisotropic electrical and optical properties,[3, 23, 24] making it promising for various device applications, including field-effect transistors,[18, 25, 26] quantum light sources,[19, 27] memristive devices,[18, 28] and polarization-sensitive photodetectors.[23, 29]

Compared to ferroelectric materials, antiferroelectric materials possess rich structural phase transition characteristics and new applications,[30] such as high-power-density energy storage,[31, 32] electromechanical actuation,[33, 34] and electrocaloric cooling.[35, 36] Although ferroelectricity has been widely investigated in various 2D ferroelectric materials as mentioned above, intrinsic 2D antiferroelectricity is much rarer than 2D ferroelectricity,[37, 38] particularly that can stably exist at room temperature. Recently, several approaches have been proposed to achieve antiferroelectricity or paraelectricity in 2D $NbOX_2$ via external stimuli, including strain field,[39, 40] temperature field,[41] and optical field.[42] For example, Mao et al. predicted the emergence of room-temperature antiferroelectricity in 2D $NbOCl_2$ under shear strain, according to molecular dynamics simulations.[40] Additionally, Huang et al. reported antiferroelectricity in few-layer $NbOCl_2$ via temperature-induced phase transition.[41] However, once these external energy fields are removed, 2D $NbOCl_2$ recovers to the ferroelectric phase. To the best of our knowledge, stable room-temperature intrinsic antiferroelectricity in 2D $NbOX_2$ has not yet been experimentally discovered.

In this work, we report the observation of room-temperature stable antiferroelectricity in layered $NbOCl_2$, and demonstrate reversible, external-field-driven antiferroelectric to ferroelectric (AFE-to-FE) phase transition. The results show that as-grown $NbOCl_2$ flakes, ranging from few-layer to bulk-like samples, all exhibit a stripe-like second-harmonic generation (SHG) distribution, indicative of pronounced polar heterogeneity across the crystal lattice. Combined incident-angle-dependent SHG and Raman spectroscopy analyses confirm that this periodic SHG distribution is not attributed to in-plane or out-of-plane ferroelectric polarization; rather, it arises from the existence of antiferroelectric phase, and/or the coexistence of antiferroelectric and ferroelectric phases. Furthermore, we achieve precise control over the AFE-to-FE phase transition by applying external stimuli along $NbOCl_2$'s polar axis. Specifically, localized strain engineering via single silver nanowire triggers spatially selective AFE-to-FE conversion. In addition, an in-plane electric field introduces a reversible SHG response, consistent with switching from the antiferroelectric state (negligible SHG) to the ferroelectric state (strong SHG). Thus, our study not only establishes $NbOCl_2$ as a reliable room-temperature 2D antiferroelectric material, but also provides a new route for manipulating its nonlinear optical responses via external-field-induced AFE-to-FE phase transition.

## 2 Results and Discussion

**Observation of unusual nonlinear optical responses in 2D $NbOCl_2$.** 2D layered $NbOX_2$ (X = Cl, I) exhibits three distinct phase structures: ferroelectric, antiferroelectric, and paraelectric.[17] Previous studies indicate that $NbOI_2$ predominantly adopts ferroelectric and paraelectric phases, whereas $NbOCl_2$ stabilizes in ferroelectric and antiferroelectric ordered states.[43] The crystal structure of $NbOCl_2$ is composed of a 2D network of $[NbO_2Cl_4]$ octahedra (**Figure 1**a), where off-center Nb displacements along the *b*-axis drive the emergence of in-plane ferroelectric (*C*2) or antiferroelectric ($C2/c$) order. In the ferroelectric phase (left panel), aligned Nb displacements break inversion symmetry and generate a net spontaneous polarization along *b*-axis. In contrast, antiparallel Nb displacements in the antiferroelectric phase (right panel) cancel the local dipoles, resulting in vanishing net polarization.[17] In this work, high-quality bulk $NbOCl_2$ crystals were synthesized via a chemical vapor transport method (Figure S1). As-grown crystals exhibit a lustrous black appearance and pronounced needle-like morphology (Figure 1b).

Owing to weak vdW interlayer interactions, $NbOCl_2$ flakes of controlled thickness can be easily achieved by mechanical exfoliation. The inset of Figure 1c shows the optical image of a representative exfoliated $NbOCl_2$ flake transferred onto a $SiO_2$/Si substrate. Raman spectroscopy reveals five characteristic vibrational modes at ~161, ~176, ~298, ~340, and ~670 $cm^{-1}$ (labeled P1-P5), consistent with the previously reported results.[19] Figures 1d and S2 show the polar plots of parallel polarized Raman peak intensity as a function of sample rotation angle, enabling unambiguous determination of the in-plane *b*-axis and *c*-axis.[23] Figure 1e presents the atomic force microscopy (AFM) image of the same flake shown in Figure 1c inset, confirming a uniform thickness of approximately 131 nm. Consistent with its high structural homogeneity, the Raman peak intensity mapping at P2 (Figure 1f) exhibits spatially uniform signal distribution, corroborating both high crystallinity and compositional homogeneity in the as-grown material.

SHG serves as a sensitive probe of non-centrosymmetry and is widely employed to characterize ferroelectric order in 2D materials.[44] Here, we focused a picosecond laser onto the sample surface at normal incidence and collected the SHG signal in reflection mode. Figure 1g displays the SHG mapping of the same flake in Figure 1e. Despite the morphological and thickness uniformity of the flake, the SHG mapping reveals a distinct stripe-like intensity distribution. Line-profile analysis demonstrates periodic oscillations of SHG signal along the *c*-axis (Figure 1h), with an average period of ~ 1.5 μm, while the signal varies smoothly and monotonically along the *b*-axis (Figure 1i). This stripe pattern persists across $NbOCl_2$ flakes spanning a broad thickness range (few-layer to bulk-like, Figure S3), underscoring the robustness and universality of the underlying periodic SHG distribution.

To elucidate the origin of this unusual spatial SHG distribution, we propose three polarization models (Figure 1j). Model 1 assumes a homogeneous, purely in-plane ferroelectric polarization, which predicts uniform and strong SHG intensity, thus inconsistent with observed spatial distribution. Model 2 involves coexisting in-plane and out-of-plane ferroelectric domains, where the out-of-plane component contributes negligibly to SHG, while the mixed regions generate intermediate SHG intensity. Model 3 instead proposes a phase mixture of ferroelectric and antiferroelectric domains: antiferroelectric regions are SHG-inactive, whereas coexisting regions produce intermediate SHG. Both Models 2 and 3 qualitatively reproduce the experimentally observed spatial SHG contrast; however, they arise from different physical origins. To evaluate these scenarios, additional experiments are designed in later sections.

**Incident angle dependent SHG measurements on 2D $NbOCl_2$.** To test Model 2, which posits the coexistence of in-plane and out-of-plane ferroelectric polarization components in 2D $NbOCl_2$, we conducted incident angle ($\varphi$) dependent SHG measurements, as shown in **Figure 2**a. Here, $\varphi$ denotes the angle between the incident light propagation direction and the surface normal of the sample. According to Model 2, at normal incidence ($\varphi$ = 0°), the SHG signal from in-plane polarization component should be maximized, whereas that from the out-of-plane component should vanish. As $\varphi$ increases, the projected electric field component along the out-of-plane direction grows, thereby enhancing the SHG contribution from out-of-plane polarization; conversely, the in-plane projection weakens, leading to a reduction in the in-plane SHG response. Consequently, the two polarization domains are expected to exhibit opposite $\varphi$-dependent SHG behavior.

Figure 2b shows the optical image of an exfoliated $NbOCl_2$ flake transferred onto a $SiO_2$/Si substrate. The corresponding SHG mapping acquired at $\varphi$ = 0° is displayed in Figure 2c. Two spatially distinct regions are clearly resolved: a low-SHG region (Region 1) and a high-SHG region (Region 2). Figure 2d compares the SHG spectra measured at representative locations within Region 1 (S1), Region 2 (S2), and the bare substrate (S3). Based on Model 2, Region 1 and Region 2 are hypothesized to be predominantly governed by out-of-plane and in-plane ferroelectric polarizations, respectively. To experimentally assess this hypothesis, we performed line-scan SHG measurement across three regions (indicated by the red solid line in Figure 2c) while continuously varying $\varphi$ from $0^{o}$ to $24^{o}$. Figure 2e presents the resulting spatial-angular SHG intensity profile, from which we extracted the $\varphi$-dependent SHG intensities at S1, S2, and S3 (Figure 2f). Critically, Region 1 exhibits negligible SHG intensity across the entire angular range, with no observable enhancement as $\varphi$ increases, whereas Region 2 displays a consistent, monotonic decrease in SHG intensity with increasing $\varphi$. These observations confirm that Region 2 is indeed dominated by in-plane polarization; however, the absence of angular enhancement in Region 1 contradicts the predicted behavior for an out-of-plane polarized domain. Therefore, the angle-resolved SHG measurements do not support Model 2. Since the SHG signal of Region 2 also drops by about two orders of magnitude between 0° and 24°, part of the angular dependence is likely instrumental. The more direct argument against Model 2 is given by the Raman spectra below, which show that the SHG-silent regions differ structurally from the ferroelectric ones.

**Evidence of intrinsic antiferroelectricity in 2D $NbOCl_2$.** To test Model 3, which posits the coexistence of ferroelectric and antiferroelectric phases in $NbOCl_2$, we conducted SHG mapping coupled with spatially resolved Raman spectroscopy. Huang et al. previously reported temperature-dependent Raman spectra of few-layer $NbOCl_2$ over the range of 300-500K, and observed a discontinuous downshift of the P1 mode from 161 to 156 $cm^{-1}$ upon heating, which was attributed to a thermally driven ferroelectric-to-antiferroelectric phase transition.[41] The high-frequency $P1_H$ mode (161 $cm^{-1}$) and low-frequency $P1_L$ mode (156 $cm^{-1}$) are considered to be contributed by the ferroelectric and antiferroelectric phases, respectively.[39, 41] A P1 Raman mode at 156 $cm^{-1}$ has been reported for $NbOCl_2$ flakes after being placed in the air for several weeks, while it was not linked to antiferroelectric order.[45] Based on these previous studies, P1 is sensitive to the polarization arraignment and can serve as spectroscopic criterion for distinguishing between ferroelectric and antiferroelectric phases.

According to Model 3, an increase in SHG intensity at room temperature should correlate with a relative enhancement of the $P1_H$ component and suppression of the $P1_L$ component, reflecting transition from the antiferroelectric to the ferroelectric phase. To test this prediction, we prepared an exfoliated $NbOCl_2$ flake on a $SiO_2$/Si substrate (Figure S4), which exhibits a stripe-like, spatially heterogeneous SHG response (**Figure 3**a). Figure 3b displays the SHG spectra acquired from seven representative spatial locations (labeled as S1-S7) across the same flake. SHG intensity decreases monotonically from maximum at S1 to a near-zero at S7. Figure 3c shows the corresponding Raman spectra acquired from these local points, using a 532-nm laser with a power of 2 mW as excitation source. As expected, as SHG intensity increases (moving from S7 to S1), the P1 peak transitions abruptly from 155.5 ($P1_L$ mode) to 161.2 $cm^{-1}$ ($P1_H$ mode). Laser power-dependent Raman measurements (Figure S5) show that the intensity ratio $I(P1_H)/I(P1_L)$ measured at a fixed spatial location remains invariant when the power is varied from 0.1 to 3 mW, thereby ruling out laser-induced heating effect. Peak-fitting analysis (Figure S4) indicates that S7 is predominately antiferroelectric ($P1_L$ fraction ~87%), whereas S1 is dominantly ferroelectric ($P1_H$ fraction ~93%). Crucially, other positions (S2-S6) exhibit simultaneous contributions from both $P1_L$ and $P1_H$ modes, providing direct experimental evidence for coexistence of the two phases. These findings not only robustly support Model 3 as the physical origin of the observed SHG contrast, but also demonstrate that antiferroelectric order in $NbOCl_2$ is intrinsically stable at room temperature, without requiring external stimuli such as electric field or strain.

To quantitatively establish the correlation between SHG response and local lattice symmetry, we performed Raman mapping across the same flake. Figures 3d and 3e present spatial mapping of the intensities for the $P1_H$ and $P1_L$ bands, respectively. From these, we computed the intensity ratio $I(P1_H)/I(P1_L)$, whose spatial distribution is shown in Figure 3f. A direct comparison of the SHG mapping (Figure 3a) and Raman mappings (Figure 3d-3f) reveals strong spatial correspondence: the stripe-like SHG contrast precisely mirrors the spatial evolution of the P1 mode components. This result indicates that the heterogeneous SHG response arises from intrinsic, spatially varying lattice distortions associated with local phase heterogeneity. To further validate this correlation, we performed a statistical analysis of SHG intensity versus $I(P1_H)/I(P1_L)$ across most of the mapped points (Figure 3g). A closely linear relationship emerges, where SHG intensity increases with increasing $I(P1_H)/I(P1_L)$, consistent with a local transition from antiferroelectric to ferroelectric order. This quantitative, positive correlation not only confirms a fundamental coupling between SHG efficiency and Raman-active P1 mode, but establishes both SHG microscopy and Raman spectroscopy as noninvasive, label-free optical probes for monitoring AFE-to-FE phase transition in $NbOCl_2$.

**Strain-induced AFE-to-FE phase transition in 2D $NbOCl_2$.** Previous studies have shown that the ferroelectric to antiferroelectric phase transition in $NbOCl_2$ can be effectively tuned by applying strain along its polar axis.[39] Building on this foundation, we propose a strategy to induce "local" AFE-to-FE phase transition in 2D $NbOCl_2$ by designing a 2D/1D hybrid structure. **Figure 4**a shows the schematic of an antiferroelectric $NbOCl_2$ layer integrated with a single silver nanowire (AgNW). The AgNW serves as a localized mechanical template, generating nonuniform tensile strain in the overlaid $NbOCl_2$ layer along the *b*-axis. This strain perturbs the relative displacement of Nb and O atoms, thereby destabilizing the antiparallel dipole arrangement characteristic of the antiferroelectric phase and promoting a net polar order consistent with the ferroelectric phase (Figure 4b).

To validate this design, we first prepared a multilayer $NbOCl_2$ flake, predominantly in the antiferroelectric phase, on a gel-film (Figure S6), followed by dry transfer onto a pre-selected AgNW on $SiO_2$/Si substrate (Figure 4c). Parallel polarized Raman measurements (Figure 4c inset) confirm that the polar *b*-axis of $NbOCl_2$ is orientated perpendicular to the nanowire axis. As shown in Figure 4b, the curvature imposed during the transfer stretches the $NbOCl_2$ layer vertically above the AgNW, resulting in tensile strain predominantly aligned along the *b*-axis. The topography image of the 2D/1D hybrid is characterized by AFM measurements, as shown in Figures 4d and 4e, where the $NbOCl_2$ flake exhibits a uniform thickness of ~89 nm, while the underlying AgNW has a diameter of ~56 nm. By applying the thin-film bending model,[45] the maximum tensile strain $\varepsilon$ at the apex of the bent $NbOCl_2$ layer is estimated using Equation (1):

$$\varepsilon = \frac{\pi^2 h t}{(1-v^2)w^2} \quad (1)$$

where $h$ is the thickness of $NbOCl_2$ (~89 nm), $t$ is the height of the curved $NbOCl_2$ (~92 nm), $w$ is the width of the curved $NbOCl_2$ (5.7 μm), and $v$ is $NbOCl_2$'s Poisson's ratio (0.24). Substituting these values yields $\varepsilon \approx 0.26\%$. It should be noted that this is an upper estimate. In an ~89 nm thick flake, the bending strain changes sign across the thickness, so the average strain probed is smaller.

We next probe the local polarization response of the strained $NbOCl_2$ using SHG microscopy. Figure 4f presents the SHG image corresponding to the sample in Figure 4c, with the $NbOCl_2$ partially coupled to the AgNW (indicated by the black dotted line). Within the unstrained regions of the $NbOCl_2$, the SHG emission is nearly undetectable, consistent with antiferroelectric symmetry. In contrast, the $NbOCl_2$ regions directly overlaid with the AgNW exhibit a pronounced SHG signal. Quantitative line profiles (Figure 4g) further corroborate this trend: SHG intensity increases sharply at the hybrid, whereas the bare AgNW, the $SiO_2$/Si substrate, and the uncovered $NbOCl_2$ regions yield negligible signals, consistent with a strain-induced AFE-to-FE transition in $NbOCl_2$, although a contribution from near-field enhancement of a residual ferroelectric signal by the AgNW cannot be fully excluded.

To further understand AgNW-induced phase transition, we acquired the SHG spectra at three representative positions (Figure 4h): S1 ($SiO_2$/Si substrate), S2 (bare $NbOCl_2$), and S3 ($NbOCl_2$/AgNW hybrid). Only S3 exhibits a measurable SHG response, with spectral characteristics matching those expected for the ferroelectric phase. Moreover, the inset of Figure 4h displays the angular dependence of parallel-polarized SHG intensity at S3. A clear two-fold polarization pattern is observed, with maxima aligned along the *b*-axis direction, providing direct evidence that the SHG signal arises from strain-induced emergence of a macroscopic polar axis in $NbOCl_2$, i.e., an AFE-to-FE phase transition.

**Electric-field-induced reversible AFE-to-FE phase transition in 2D $NbOCl_2$.** Theoretical predictions indicate that an external electric field can drive a phase transition from the antiferroelectric to the ferroelectric state in $NbOCl_2$.[40] Motivated by this prediction, we experimentally investigate the electric-field-controlled reversible AFE-to-FE phase transition. **Figure 5**a presents a schematic of the two-terminal antiferroelectric device: a bias voltage ($V_{bias}$) is applied across the channel to generate a tunable electric field along the *b*-axis of $NbOCl_2$, while SHG signals are collected *in-situ* from the channel region to monitor the evolution of structural symmetry during the transition. Figure 5b shows the optical image of a representative multilayer $NbOCl_2$ device, with a channel length of ~5 μm and its *b*-axis oriented parallel to the current path, as confirmed by parallel polarized Raman measurement (inset). The corresponding SHG mapping acquired at $V_{bias}$ = 0 V is shown in Figure 5c. The line-profile analysis shows a near-zero SHG intensity across the channel, confirming an initial centrosymmetric antiferroelectric-dominated state.

Figure 5d displays the $V_{bias}$-dependent SHG spectra recorded during voltage sweeps up to a maximum voltage ($V_{max}$) of -21V, where arrowheads indicate the sweep direction. As $V_{bias}$ is swept from 0 V to -21 V (left panel), the SHG signal emerges gradually and then increases sharply. Upon reversal of the sweep from -21 V to 0 V (right panel), the SHG intensity remains elevated initially before rapidly decaying to near-zero levels. This pronounced hysteresis in SHG intensity, which is dependent on both $V_{bias}$ magnitude and sweep direction, is consistent with a reversible, electric-field-driven AFE-to-FE phase transition in 2D $NbOCl_2$.

We further examine the SHG intensity-$V_{bias}$ hysteresis loops measured at varying $|V_{max}|$ (Figure 5e). No measurable SHG emission is observed for $|V_{max}|$ < 10 V. For $|V_{max}|$ > 10 V, distinct hysteretic SHG response emerges exclusively under negative bias sweeps, whereas no detectable SHG modulation occurs under positive bias. This unipolar response differs from the symmetric double hysteresis loop expected for a purely field-driven AFE-to-FE transition. A likely origin is an asymmetry of the two graphene/$NbOCl_2$ contacts, which concentrates the field near one electrode. A contribution from bias-driven ion migration, which underlies the memristive switching reported for $NbOCl_2$,[28] cannot be excluded at present. By extracting the maximum SHG intensity ($I_{max}$) for both bias directions (Figure 5f), we find a robust positive correlation between $I_{max}$ and $V_{max}$ when $V_{max}$ < -10 V. These results indicate a threshold switching voltage of ~10 V, corresponding to a critical AFE-to-FE field of ~20 kV/cm, required to overcome the energy barrier of transition from antiferroelectric phase (free of SHG response) to ferroelectric phase (strong SHG intensity). These findings point to 2D antiferroelectric $NbOCl_2$ as a promising platform for ultralow-power optoelectronic devices capable of electric-field-programmable nonlinear optical readout, which enables novel “write-by-voltage, read-by-SHG” logic operation.

## 3 Conclusion

In summary, we have successfully observed intrinsic antiferroelectricity in layered $NbOCl_2$ at room temperature and achieved precise control over the AFE-to-FE phase transition by applying external stimuli. We find that as-grown $NbOCl_2$ flakes, ranging from few-layer to bulk-like samples, all exhibit a distinct stripe-like SHG spatial distribution. A combination of incident-angle-dependent SHG and Raman spectroscopy analyses demonstrates that this unique SHG behavior is not attributed to in-plane or out-of-plane ferroelectric polarization, but originates from the existence of antiferroelectric phase and the coexistence of antiferroelectric and ferroelectric phases. In addition, we have developed a local strain engineering strategy to induce spatially selective AFE-to-FE phase transition in 2D $NbOCl_2$ by integrating it with 1D Ag nanowires. Furthermore, an ultralow electric field (~20 kV/cm) applied along the *b*-axis induces a reversible SHG response, consistent with switching from the antiferroelectric to the ferroelectric phase, while the origin of its polarity dependence remains to be clarified. These findings establish $NbOCl_2$ as a stable room-temperature 2D antiferroelectric material, and provide a new strategy for dynamic control of nonlinear optical responses in $NbOCl_2$ via low-power electric field induced AFE-to-FE phase transition.

## 4 Experimental Section

**Sample preparation.** Bulk $NbOCl_2$ single crystals were prepared via a chemical vapor transport method (Figure S1).[26] The precursor powders of $Nb_2O_5$, Nb, $NbCl_5$ with a molar ratio of 1:1:5 were sealed in an evacuated quartz tube. The sealed tube was placed in a two-zone furnace, heated to 700 $^oC$/450 $^oC$, kept at this temperature for six hours, and then cooled to room temperature naturally. Large-sized single crystals were collected from the cold end of the tube. 2D $NbOCl_2$ flakes with different layer thicknesses were obtained by mechanical exfoliation and then transferred onto $SiO_2$/Si substrates for further characterization. 2D/1D $NbOCl_2$/AgNW hybrids were prepared by a dry transfer method.

**Structural characterization.** The morphology and thickness of exfoliated 2D $NbOCl_2$ flakes were characterized by combining optical microscope (Eclipse LV150N, Nikon) and atomic force microscope system (Asylum Research MFP-3D, Oxford, UK) with an AFM tip (Micromesh HQ:NSC14, $k$ = 5) working at AC mode.

**Optical characterization.** The optical images of the samples were taken using Nikon Eclipse LV 150N microscope. Raman and SHG measurements were carried out at room temperature in a confocal Raman spectrometer system (WiTec Alpha 300R, Oxford instrument). Raman signals were collected by focusing a 532 nm laser beam with a power of ~2 mW onto the sample surface through a 100× objective. SHG signals were collected by focusing a 1064 nm pico-second laser beam with a power of ~10 mW onto the sample surface through a 100× objective. The SHG-silent regions identified with 1064 nm excitation coincide with the $P1_L$-rich regions in the Raman mapping recorded at 532 nm, which argues against laser heating as the origin of the $P1_L$ band. Angle-resolved Raman and SHG measurements were performed by rotating sample from $\theta$ = 0° to 360°, while maintaining the orientations of polarizer ($\overrightarrow{P_p}$) and analyzer ($\overrightarrow{P_a}$) fixed.

**Device fabrication and characterization.** Two-terminal $NbOCl_2$ devices were fabricated using standard photolithography and dry transfer technique. Firstly, patterned electrodes (5 nm Ti/25 nm Ni) were prepared on $SiO_2$/Si substrate via photolithography (Smart print UV, microlight 3D), followed by transferring few-layer graphene on top as an intermediate contact layer. Secondly, multilayer $NbOCl_2$ flake was mechanically exfoliated on gel-film from its bulk crystal. Thirdly, the $NbOCl_2$ flake were precisely transferred onto the graphene electrodes. During the transfer process, the *b*-axis of $NbOCl_2$ was parallel to channel direction. Electric-field-induced phase transition experiments were conducted by a combined semiconductor parameter analyzer (B1500A, Keysight) and SHG spectrometer system.

## Conflict of Interest

The authors declare no competing interest.

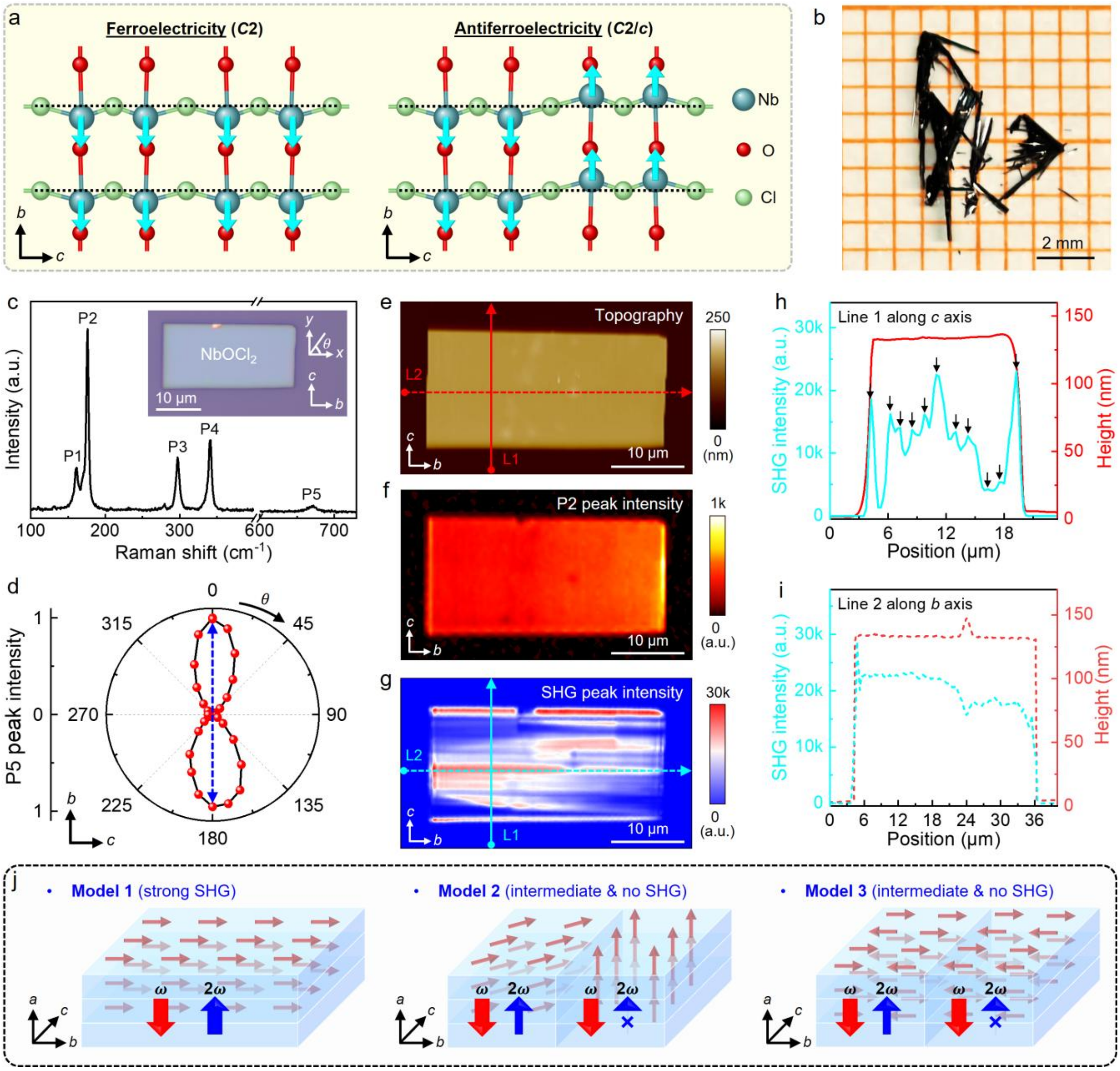

**Figure 1.** Observation of unusual nonlinear optical responses in 2D $NbOCl_2$. (a) Crystal structures of monolayer ferroelectric ($C2$) and antiferroelectric ($C2/c$) $NbOCl_2$. Arrows mark the Nb displacements. (b) Optical photograph of as-grown single-crystalline $NbOCl_2$ bulk samples. (c) Raman spectrum acquired from an exfoliated $NbOCl_2$ flake on $SiO_2$/Si substrate. Inset: the corresponding optical image and the laboratory coordinate system. (d) Polar plot of parallel-polarized Raman peak intensity for P5 mode as a function of the sample rotation angle $\theta$. (e) AFM topography image, (f) P2 Raman peak intensity mapping, and (g) SHG intensity mapping of the same sample in (c) inset. (h) SHG intensity and height profiles along the solid line 1 in (e) and (g). (i) SHG intensity and height profiles along dashed line 2 in (e) and (g). (j) The proposed polarization models to explain unusual SHG responses in 2D $NbOCl_2$.

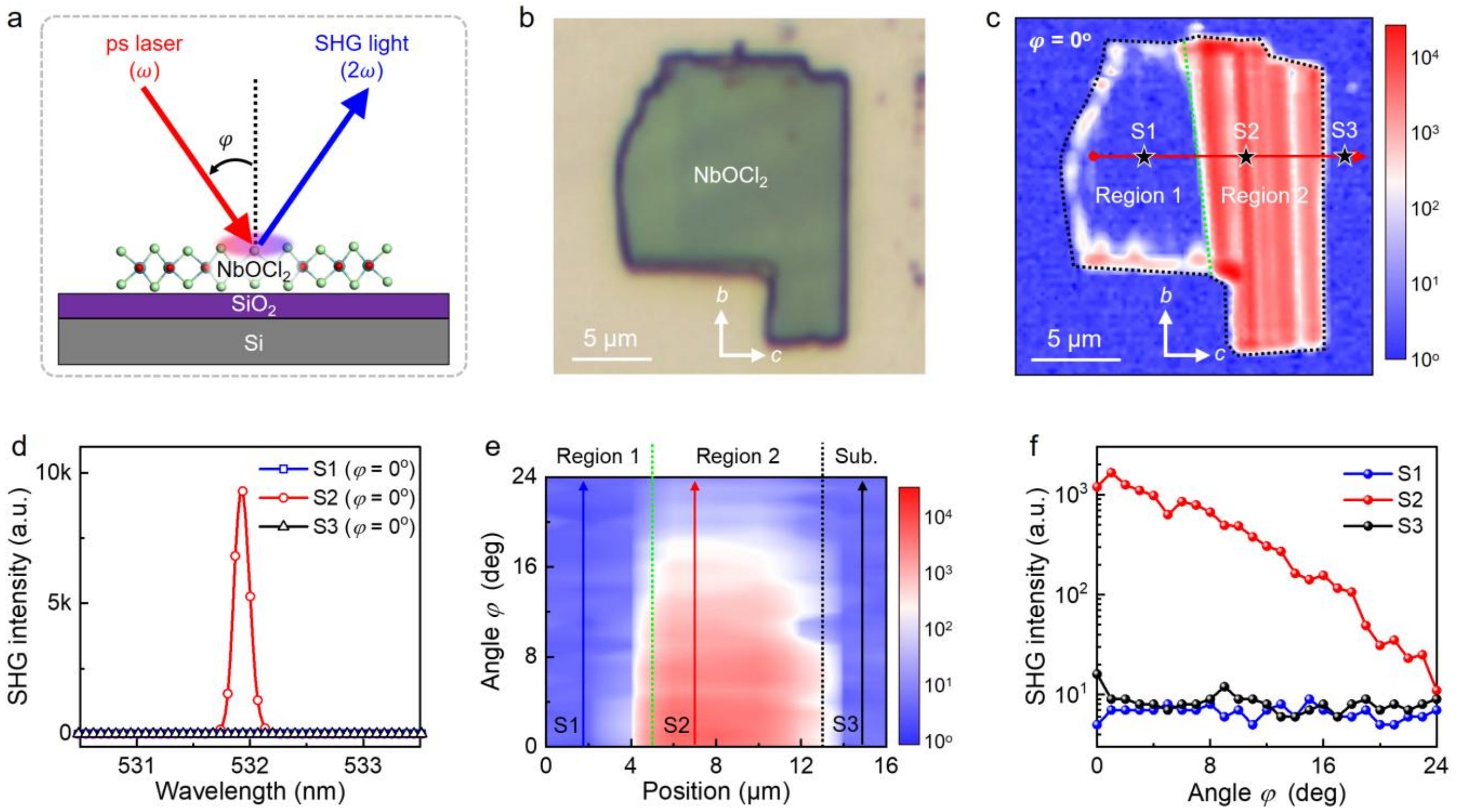

**Figure 2.** Incident angle dependent SHG measurements on 2D $NbOCl_2$. (a) Schematic of the experimental setup for incident angle dependent SHG measurements. (b) Optical image of an exfoliated $NbOCl_2$ flake, with (c) the corresponding SHG mapping taken at incident angle $\varphi$ of 0°. (d) SHG spectra collected from three local regions in panel (c). (e) SHG intensity profile along the red line 1 in (c) as a function of the incident angle $\varphi$. (f) The dependence of SHG intensity at S1 (Region 1), S2 (Region 2), and S3 (Substrate Region) in (e) on incident angle $\varphi$.

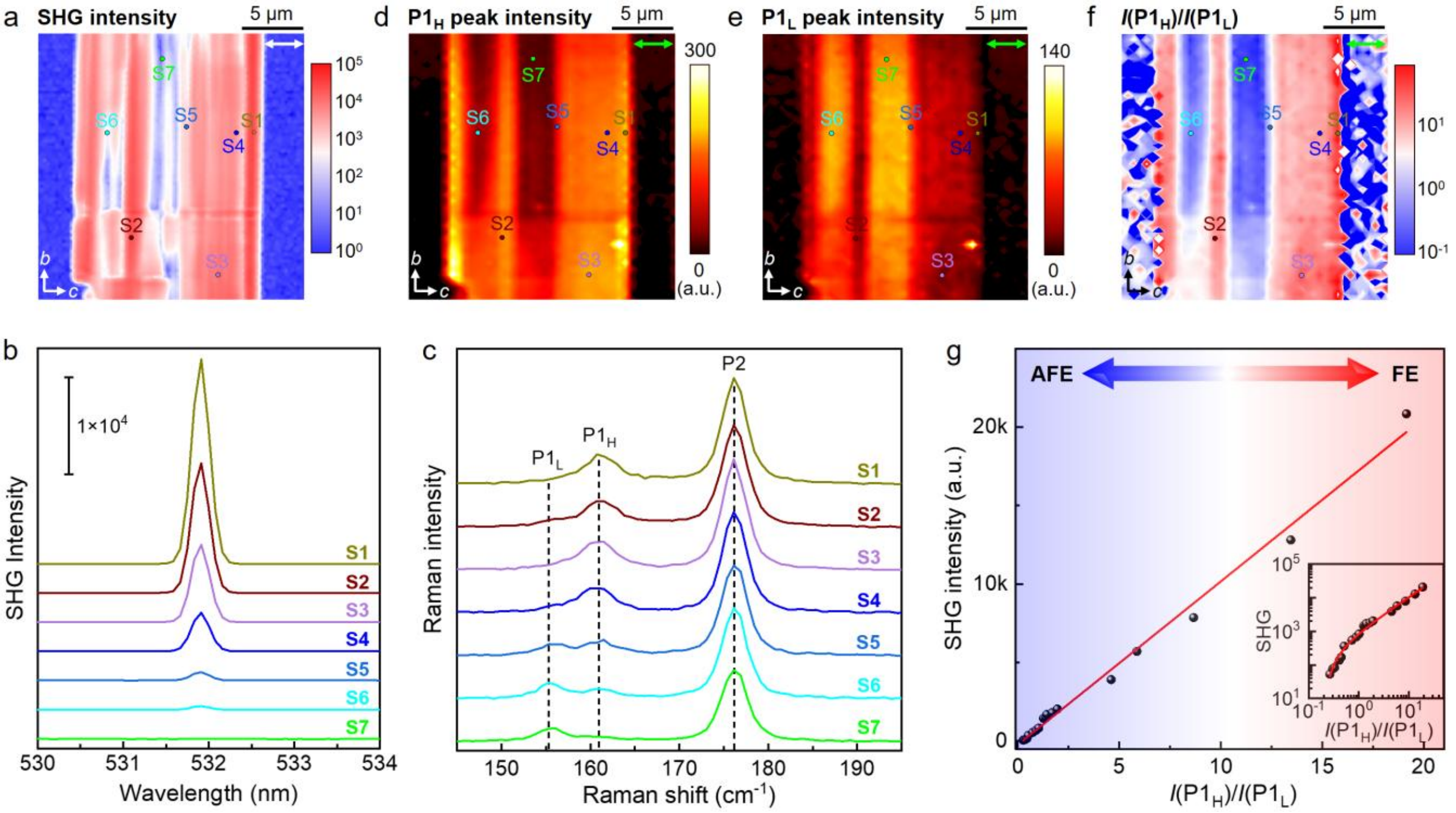

**Figure 3.** Evidence of intrinsic antiferroelectricity in 2D $NbOCl_2$. (a) SHG intensity mapping for an exfoliated $NbOCl_2$ flake on $SiO_2$/Si substrate. (b) SHG spectra and (c) Raman spectra acquired from seven local points (S1, S2, S3, S4, S5, S6, and S7) in (a). (d-e) Raman peak intensity mapping at (d) $P1_H$ band and (e) $P1_L$ band for the same sample in (a). (f) Raman mapping of peak intensity ratio [$I(P1_H)/I(P1_L)$] extracted from (d) and (e). (g) Correlation between SHG intensity and $I(P1_H)/I(P1_L)$ ratio in 2D $NbOCl_2$.

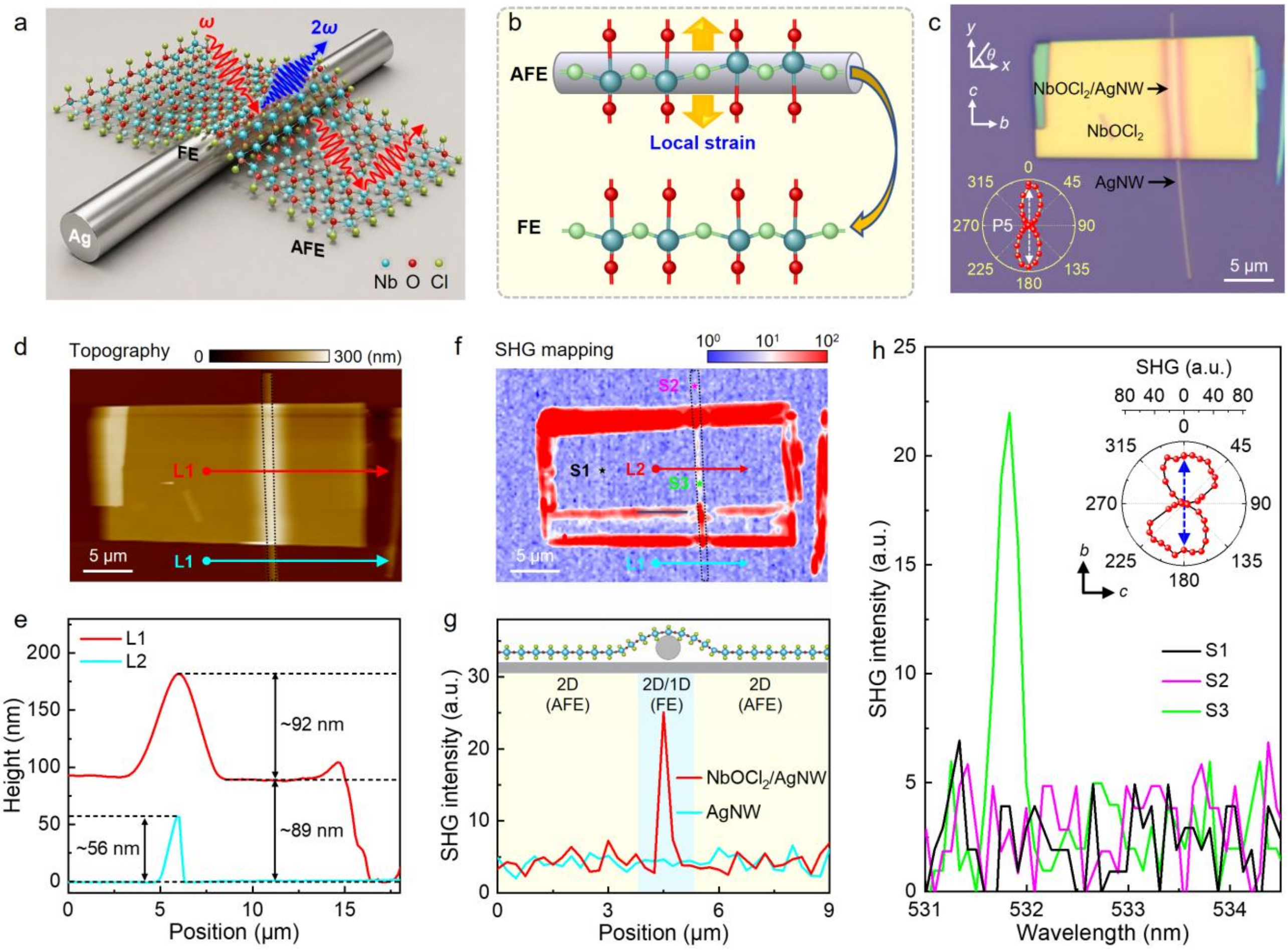

**Figure 4.** Strain-induced AFE-to-FE phase transition in 2D $NbOCl_2$. (a) Schematic of an antiferroelectric $NbOCl_2$ layer integrated with single AgNW. (b) Schematic of local strain formed in $NbOCl_2$ layer, leading to AFE-to-FE phase transition. (c) Optical image of a 2D/1D $NbOCl_2$/AgNW hybrid on $SiO_2$/Si substrate. The *b*-axis of $NbOCl_2$ is perpendicular to AgNW, as evidenced by parallel-polarized Raman analysis (inset). (d) AFM topography of the same sample in (c), with (e) the corresponding height profiles along two lines in panel (d). (f) SHG intensity mapping of the same sample in (d), with (g) the corresponding SHG intensity profiles along two lines in panel (f). (h) SHG spectra for three local points (S1, S2, and S3) in panel (f). Inset: polar plot of parallel polarized SHG intensity at $NbOCl_2$/AgNW hybrid as a function of the sample rotation angle $\theta$.

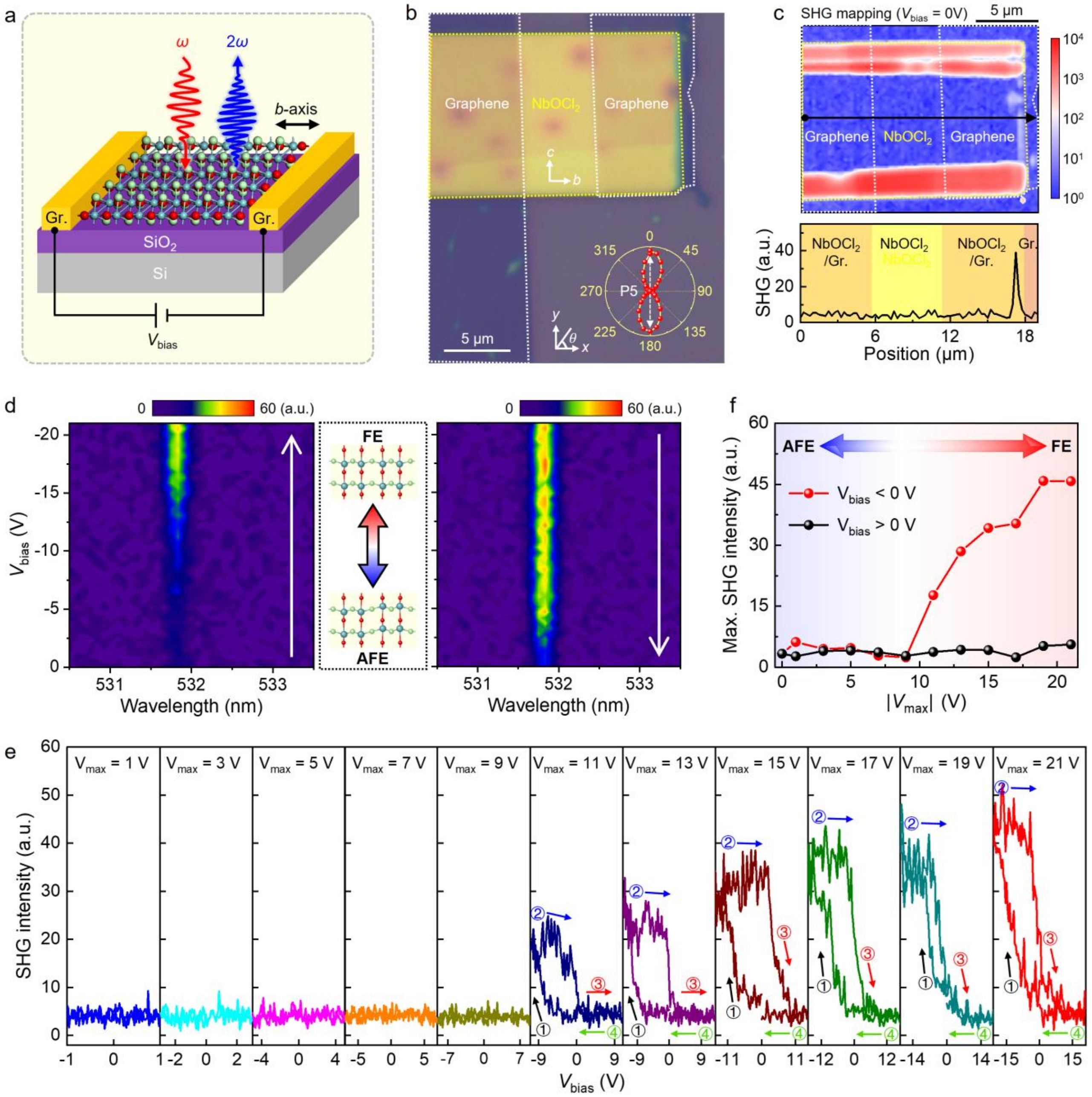

**Figure 5.** Electric-field-induced reversible AFE-to-FE phase transition in 2D $NbOCl_2$. (a) Schematic of the two-terminal $NbOCl_2$ device used for electric-field-driven phase transition. (b) Optical image of a 2D $NbOCl_2$ device, with its *b*-axis parallel to the channel direction, as evidenced by parallel polarized Raman analysis (inset). (c) SHG intensity mapping of the same device in (b) taken at $V_{bias}$ = 0 V, with the SHG intensity profile along the solid line (inset). (d) SHG spectra of the $NbOCl_2$ device in the channel region by sweeping $V_{bias}$ (left) from 0 to -21V and then (right) from -21 to 0V, indicating reversible AFE-FE phase transition. (e) SHG intensity-$V_{bias}$ curves of the $NbOCl_2$ device taken at different $V_{max}$. The arrows illustrate the switching direction. (f) Relationship between maximum SHG intensity and $V_{max}$ extracted from (e).